\documentstyle[12pt]{article}

\begin{document}
\begin{titlepage}
\title{\bf $2+1$ KdV(N) Equations}
\vspace{5cm}
\author {
Metin G{\" u}rses\thanks{Email:gurses@fen.bilkent.edu.tr} \ and Asl{\i} Pekcan\thanks{Email:asli@fen.bilkent.edu.tr}\\
{\small Department of Mathematics, Faculty of Science} \\
{\small Bilkent University, 06800 Ankara - Turkey}}

 \maketitle
\begin{abstract}
We present some nonlinear partial differential equations in
$2+1$-dimensions derived from the KdV Equation and its symmetries.
We show that all these equations have the same 3-soliton
structures. The only difference in these solutions are  the
dispersion relations. We also show that they posses the Painlev{\'
e} property.
\end{abstract}
\end{titlepage}

\section{Introduction}
After Karasu et al \cite{kar1} and Kupershmidt's \cite{kup} works
there have been some attempts to enlarge  classes of such
integrable  nonlinear partial differential equations, known as the
hierarchy of KdV(6) equations, and to obtain their
$2+1$-dimensional extensions \cite{kun1}-\cite{her}. These
equations are not in local evolutionary form. Due to this reason
their integrability is examined by studying their Painlev{\' e}
property and by the existence of soliton solutions by Hirota
method rather then searching for recursion operators.

Although they differ in some examples \cite{sak}, the Painlev{\'
e} property and the Hirota bilinear approach are powerful tests to
examine integrability. In particular, existence of 3-soliton
solutions of a nonlinear partial differential equation is believed
to be an important indication for the integrability
\cite{hirota}-\cite{hiet2}. Using this conjecture as a test of
integrability, we propose $2+1$-dimensional
generalizations of the KdV(N) equations and present their
3-soliton solutions.

In this work we first give these equations in general. Such
equations exist not only for KdV family but also for all
integrable equations with recursion operators. Concentrating on KdV(N) type of equations, we propose their $2+1$
extensions and give 3-soliton solutions of these proposed
equations. We then show that all these equations possess the
Painlev{\' e} property.

Let $u_{t}= F[u]$ be a system of  integrable nonlinear partial
differential equations, where $F$ is a function of
$u,u_{x},u_{xx}, \cdots$. Let $\sigma_{n}, (n=0,1,2,\cdots)$ be
infinite number of commuting symmetries. One can write these
symmetries as $\sigma_{n}=R^n\, \sigma_{0}$, where $\sigma_{0}$ is
one of the symmetries of the equation and $R$ is the recursion
operator. Then the corresponding evolution equations are given as

\begin{equation}\label{kdvn}
u_{t_{n}}=R^n\, \sigma_{0}, ~~ n=0,1,2, \cdots.
\end{equation}
For $n=1,2,3, \cdots$, Eq.(\ref{kdvn}) produces hierarchy of the
equation $u_{t}= F[u]$. Since superposition of symmetries is also
a symmetry the above equations can be extended to the following
more general type

\begin{equation}
u_{t_{nm}}=a\,R^n\, \sigma_{0}+ b\,R^m\, \sigma_{1}, ~~ m,n=0,1,2,
\cdots,
\end{equation}
where $\sigma_{1}$ is another symmetry, $a$ and $b$ are arbitrary
constants.

Interesting classes of equations are obtained by letting $m$ as a
negative integer. As an example letting  $m=-1$ we get

\begin{equation}\label{non}
u_{t_{n}}=a\,R^n\, \sigma_{0}+ b\,R^{-1}(\sigma_{1}), ~~ n=0,1,2,
\cdots .
\end{equation}
This equation is in evolutionary type but nonlocal, because
$R^{-1}(\sigma_{1})$ term is an infinite sum of terms containing
$D^{-1}$. It is possible to write (\ref{non}) as a local
differential equation by multiplying both sides by the recursion
operator $R$ which is given by
\begin{equation}\label{son}
R\,[u_{t_{n}}-a\,R^n\, \sigma_{0}]=b\, \sigma_{1}, ~~ n=0,1,2,
\cdots .
\end{equation}
For some integrable equations $R(0)$ may not vanish. For example
it is proportional to $u_{x}$ for the KdV equation. For this
reason the constants $a$ and $b$ are introduced for convenience.
The above classes of equations (\ref{son}) are our basic starting
point in this work. For a given integrable equation $u_{t}= F[u]$
it is expected that all the above equations are also integrable.
Karasu et al's \cite{kar1} work corresponds to $b=0$, $a=-1, n=1$
for the KdV equation. This equation and its higher order versions
KdV($2n+4$), $n=2,3,\cdots$ are also integrable \cite{wang}. These
equations are all in $1+1$-dimensions.

\section{2+1 KdV(2n+4) Family and 3-Soliton Solutions}

Taking the original equation as the KdV equation, we will
investigate 3-soliton solutions of the classes of equations in
(\ref{son}) in $2+1$-dimensions corresponding to different values
of ($a,b,n,\sigma_{0}, \sigma_{1}$). We conjecture that all these
equations are integrable in some sense. Here we examine the
integrability by the existence of the 3-soliton solutions. The
well-known KdV equation is given as
\begin{equation}
u_{t}+u_{xxx}+12\,u\,u_{x}=0,
\end{equation}
with the recursion operator
\begin{equation}
R=D^2+8\,u+4\,u_{x}\,D^{-1}.
\end{equation}
For certain values of the set ($a,b,n,\sigma_{0}, \sigma_{1}$) we
show that the corresponding  $2+1$-dimensional equations possess
the same 3-soliton structures as of the KdV equation and its
hierarchy except their dispersion relations.

We obtain $2+1$-dimensional equations by assuming $u=u(x,t,y)$
where $y$ is a new independent variable and by letting one of the
symmetries $ \sigma_{0}=u_{y}$ and  ($a=-1,b=0,n=1$). Then we get
\begin{equation}
u_{t}+u_{xxy}+8u\,u_{y}+4u_{x}D^{-1}\,u_{y}=0.
\end{equation}
By letting $u=v_{x}$ we get a local equation as
\begin{equation}
v_{tx}+v_{xxxy}+8v_{x}\,v_{xy}+4v_{xx}\,v_{y}=0.
\end{equation}
This equation has 3-soliton solutions. Let $v=f_{x}/f$ then
\begin{equation}
f=1+f_{1}+f_{2}+f_{3}+A_{12}\,f_{1}\,f_{2}+A_{13}\,f_{1}\,f_{3}+A_{23}\,f_{2}\,f_{3}+A_{123}\,f_{1}\,f_{2}\,f_{3},
\end{equation}
where
\begin{eqnarray}
f_{i}&=&e^{w_{i}\,t+k_{i}\,x+\ell_{i}\,y+a_{i}},~~(i=1,2,3),\\
A_{ij}&=&\left({k_{i}-k_{j} \over k_{i}+k_{j}}
\right)^2,~~ (i,j=1,2,3),\\
A_{123}&=&A_{12}\,A_{13}\,A_{23}.
\end{eqnarray}
Here $w_{i}, k_{i}, \ell_{i}, a_{i}~~ (i=1,2,3)$ are arbitrary
constants. Dispersion relations are
\begin{equation}
w_{i}=-\ell_{i}\,k_{i}^2, ~~ (i=1,2,3).
\end{equation}

\vspace{0.3cm}

It is possible to show that all KdV($2n+4$), ($n=2,3, \cdots$)
equations $R\,(u_{t}+R^{n}\,u_{y})=0$ have 3-soliton solutions
same as the 3-soliton solutions of the KdV hierarchy in
(\ref{kdvn}).

\vspace{0.3cm}

\vspace{0.3cm} Below we give a different class of
$2+1$-dimensional equations (\ref{son}) by letting
$\sigma_{1}=u_{y}$, $\sigma_{0}=u_{x}$ and $a=b=-1$.

\vspace{0.3cm}

\noindent {\bf 1. ($n=1$) \,\,$2+1$ KdV(6) Equation:}

Eq.(\ref{son}) reduces to  $R\,(u_{t}+R\,u_{x})+u_{y}=0$. Letting
$u=v_{x}$ we get
\begin{eqnarray}
v_{t}+v_{xxx}+6v_{x}^2=q,\\
v_{xy}+q_{xxx}+8v_{x}\,q_{x}+4v_{xx}\,q=0.
\end{eqnarray}

Dispersion relations are
\begin{equation}
w_{i}=-k_{i}^3-{\ell_{i} \over k_{i}^2}, ~~~i=1,2,3.
\end{equation}

\vspace{3cm}

\noindent {\bf 2. ($n=2$)\,\, $ 2+1$ KdV(8) Equation:}

Eq.(\ref{son}) reduces to  $R\,(u_{t}+R^{2}\,u_{x})+u_{y}=0$.
Letting $u=v_{x}$ we get
\begin{eqnarray}
v_{t}+v_{xxxxx}+20\,v_{x}\,v_{xxx}+10\,v_{xx}^2+40\,v_{x}^3=q,\\
v_{xy}+q_{xxx}+8v_{x}\,q_{x}+4v_{xx}\,q=0.
\end{eqnarray}

Dispersion relations are
\begin{equation}
w_{i}=-k_{i}^5-{\ell_{i} \over k_{i}^2}, ~~~i=1,2,3.
\end{equation}

\vspace{0.3cm}

\noindent {\bf 3. ($n=3$)\,\, $ 2+1$ KdV(10) Equation:}

Eq.(\ref{son}) reduces to $R\,(u_{t}+R^{3}\,u_{x})+u_{y}=0$.
Letting $u=v_{x}$ we get
\begin{eqnarray}
v_{t}+v_{7x}+42\,v_{3x}^{2}+280\,\,v_{x}^4+56\,v_{2x}\,v_{4x}+280\,v_{x}\,v_{2x}^2 \nonumber\\
+28\,v_x\,v_{5x}+280\,v_{x}^2\,v_{3x}=q,\\
v_{xy}+q_{xxx}+8v_{x}\,q_{x}+4v_{xx}\,q=0.
\end{eqnarray}

Dispersion relations are
\begin{equation}
w_{i}=-k_{i}^7-{\ell_{i} \over k_{i}^2}, ~~~i=1,2,3.
\end{equation}

In addition to above equations  we conjecture that 3-soliton
solutions of the $2+1$ KdV($2n+4$) equation exists for all $n \ge 4$
and they are given as

\vspace{0.3cm}

\noindent {\bf 4. ($n \ge 4$)\,\, $ 2+1$ KdV($2n+4$) Equation:}
For all values of $n \ge 4$ we have
\begin{eqnarray}
u_{t}+R^{n}\,u_{x}=q,\\
v_{xy}+q_{xxx}+8v_{x}\,q_{x}+4v_{xx}\,q=0.
\end{eqnarray}
Here $u=v_{x}$. Dispersion relations are
\begin{equation}
w_{i}=-k_{i}^{2n+1}-{\ell_{i} \over k_{i}^2}, ~~~i=1,2,3.
\end{equation}

\vspace{1cm}

\section{Painlev\'{e} Property of 2+1 KdV(2n+4) Equations for n=1,2,3}

In this section, we check the Painlev{\' e} property of the $2+1$
KdV$(2n+4)$ equations for $n=1,2,3$. We used a Maple package
called PDEPtest \cite{Xu-Li} for this purpose. Here the
WTC-Kruskal algorithm is used \cite{WTCKruskal}-\cite{Kruskal}.
Note that a nonlinear partial differential equation is said to
possess the Painlev\'{e} property, if all solutions of it can be
expressed as Laurent series
\begin{equation}\label{Laurent}
v^{(i)}(x,t,y)=\sum_{j=0}^{\infty}v_j^{(i)}(x,t,y)\phi(x,t,y)^{(j+\alpha_i)}, \quad i=1,...,m,
\end{equation}
with sufficient number of arbitrary functions as the order of the equation, $v_j^{(i)}(x,t,y)$ are analytic functions, $\alpha_i$ are
negative integers.

\vspace{0.3cm} \noindent {\bf 1. $2+1$ KdV(6) Equation}.

By leading order analysis, we see that $2+1$ KdV(6) equation
admits two branches. The leading exponents for these two branches
are $-1$, and the leading order coefficients are
$$i)\quad v_0=\phi_x\quad \quad , \quad \quad ii)\quad v_0=3\phi_x.$$ The corresponding truncated expansions for these two branches are
$$i)\quad v=\frac{\phi_x}{\phi}+v_1\quad \quad , \quad \quad ii)\quad v=\frac{3\phi_x}{\phi}+v_1.$$
The resonances of the above branches are
$$i)\quad r=-1,1,2,5,6,8 \quad \quad , \quad \quad ii)\quad r=-1,1,-3,6,8,10.$$
It is clear that branch (i) is the principal(generic) one and the other one is secondary(non-generic) branch. For the principal branch, the coefficients
of the series (\ref{Laurent}) at non-resonances are
\begin{eqnarray*}
&&v_0=1, \quad v_3=0,\\
&&v_4=-\frac{1}{10}v_2\psi_t+v_2^2-\frac{1}{120}\psi_y+\frac{1}{30}v_{1_{t}},\\
&&v_7=-v_2v_5+\frac{1}{20}v_5\psi_t+\frac{1}{480}v_{2_{y}}
+\frac{1}{480}v_{2_{t}}\psi_t+\frac{1}{480}v_2\psi_{tt}\\
&&\quad\quad-\frac{1}{40}v_2v_{2_{t}}
+\frac{1}{5760}\psi_{yt}-\frac{1}{1440}v_{1_{tt}},
\end{eqnarray*}
where $v_1,v_2,v_5,v_6,v_8$ are arbitrary functions of the variables $y$ and $t$ and $\phi(x,t,y)=x-\psi(t,y)$.

\vspace{0.3cm}
\noindent
For the second branch, the coefficients of (\ref{Laurent}) at non-resonances are
\begin{eqnarray*}
&& v_0=3, \quad v_2=\frac{1}{20}\psi_t,\\
&&v_4=-\frac{1}{2800}\psi_t^2+\frac{1}{120}v_{1_{t}}-\frac{1}{840}\psi_{y},\\
&&v_5=0,\quad v_7=-\frac{1}{28800}\psi_t\psi_{tt}-\frac{1}{14400}\psi_{yt}+\frac{1}{7200}v_{1_{tt}},\\
&&v_9=-\frac{1}{806400}\psi_t\psi_{yt}+\frac{1}{201600}v_{1_{yt}}-\frac{1}{806400}\psi_{yy}-
\frac{1}{80}v_{6_{t}}\\
&&\quad\quad-\frac{1}{1680000}\psi_t^2\psi_{tt}+\frac{1}{504000}\psi_tv_{1_{tt}}+\frac{1}{252000}\psi_{tt}v_{1_{t}}
-\frac{1}{1008000}\psi_{tt}\psi_y,
\end{eqnarray*}
where $v_1,v_6,v_8,v_{10}$ are arbitrary functions of the variables $y$ and $t$ and $\phi(x,t,y)=x-\psi(t,y)$.

\vspace{0.3cm} \noindent {\bf 2. $2+1$ KdV(8) Equation}.

For the $2+1$ KdV(8) equation, we get the following information from the Painlev\'{e} property. By leading order analysis, we see that $2+1$ KdV(8) equation
admits three branches. The leading exponents for these three branches are $-1$, and the leading order coefficients are
$$i)\quad v_0=\phi_x\quad , \quad ii)\quad v_0=3\phi_x \quad , \quad iii)\quad v_0=6\phi_x.$$ The corresponding truncated expansions for these three branches are
$$i)\quad v=\frac{\phi_x}{\phi}+v_1 \quad ,  \quad ii)\quad v=\frac{3\phi_x}{\phi}+v_1\quad , \quad iii)\quad v=\frac{6\phi_x}{\phi}+v_1.$$
The resonances of the above branches are
$$i)\quad r=-1,1,2,4,5,7,8,10 \quad , \quad ii)\quad r=-1,1,2,-3,7,8,10,12\quad ,$$
$$iii)\quad r=-1,1,-3,-5,8,10,12,14.$$
 Obviously, branch (i) is the principal one and the other two are secondary branches. For the principal branch, the coefficients
of the series (\ref{Laurent}) at non-resonances are
\begin{eqnarray*}
&&v_0=1, \quad v_3=0,\\
&&v_6=-3v_2v_4-\frac{3}{280}v_2\psi_t+\frac{1}{280}v_{1_{t}}-\frac{1}{1120}\psi_y+v_2^3,\\
&&v_9=-\frac{1}{2240}v_{4_{t}}+\frac{3}{2800}v_5\psi_t+\frac{1}{22400}v_{2_{y}}-\frac{3}{5}v_2v_7\\
&&\quad\quad-\frac{3}{10}v_4v_5-\frac{3}{10}v_2^2v_5+\frac{1}{2800}v_2v_{2_t},
\end{eqnarray*}
where $v_1,v_2,v_4,v_5,v_7,v_8,v_{10}$  are arbitrary functions of the variables $y$ and $t$ and $\phi(x,t,y)=x-\psi(t,y)$ .

\vspace{0.3cm}
\noindent For the second branch, the coefficients of (\ref{Laurent}) at non-resonances are
\begin{eqnarray*}
&& v_0=3, \quad v_3=0,\\
&&v_4=\frac{1}{3}v_2^2-\frac{1}{840}\psi_t,\quad v_5=0,\\
&&v_6=\frac{2}{3}v_2^3-\frac{1}{2520}v_{1_t}-\frac{1}{630}v_2\psi_t+\frac{1}{10080}\psi_y,\\
&&v_9=-v_2v_7+\frac{1}{2016}v_2v_{2_t}-\frac{1}{806400}\psi_{tt}-\frac{1}{40320}v_{2_y},\\
&&v_{11}=\frac{1}{2520}v_7\psi_t+\frac{1}{3}v_2^2v_7+\frac{1}{45360}v_2v_{2_y}-\frac{1}{25401600}\psi_{ty}\\
&&\quad\quad -\frac{1}{2268}v_2^2v_{2_t}+\frac{1}{1411200}v_2\psi_{tt}+\frac{1}{12700800}v_{1_{tt}},
\end{eqnarray*}
where $v_1,v_2,v_7,v_8,v_{10},v_{12}$ are arbitrary functions of the variables $y$ and $t$ and $\phi(x,t,y)=x-\psi(t,y)$.

\vspace{0.3cm}
\noindent For the third branch, the coefficients of (\ref{Laurent}) at non-resonances are
\begin{eqnarray*}
&& v_0=6, \quad v_2=0, \quad v_3=0, \quad v_4=-\frac{1}{2520}\psi_t,\\
&&v_6=\frac{1}{110880}\psi_y-\frac{1}{27720}v_{1_{t}},\quad v_7=0,\\
&&v_9=\frac{1}{5644800}\psi_{tt}, \quad v_{11}=\frac{1}{101606400}\psi_{ty}-\frac{1}{50803200}v_{1_{tt}},\\
&&v_{13}=\frac{29}{384072192000}\psi_{tt}\psi_t+\frac{1}{16094453760}\psi_{yy}-\frac{1}{4023613440}v_{1_{ty}}\\
&&\quad\quad -\frac{13}{120960}v_{8_{t}},
\end{eqnarray*}
where $v_1,v_8,v_{10},v_{12},v_{14}$ are arbitrary functions of the variables $y$ and $t$ and $\phi(x,t,y)=x-\psi(t,y)$.

\vspace{0.3cm} \noindent {\bf 3. $2+1$ KdV(10) Equation}.

By leading order analysis, we see that $2+1$ KdV(10) equation
admits four branches. The leading exponents for these four branches are $-1$, and the leading order coefficients are
$$i)\quad v_0=\phi_x\quad , \quad ii)\quad v_0=3\phi_x \quad ,$$
$$ iii)\quad v_0=6\phi_x\quad ,\quad iv)v_0=10\phi_x.$$
The corresponding truncated expansions for these four branches are
$$i)\quad v=\frac{\phi_x}{\phi}+v_1 \quad ,  \quad ii)\quad v=\frac{3\phi_x}{\phi}+v_1\quad ,$$
$$ iii)\quad v=\frac{6\phi_x}{\phi}+v_1\quad , \quad iv)\quad v=\frac{10\phi_x}{\phi}+v_1.$$
The resonances of the above branches are
\begin{eqnarray*}
&&i)\quad r=-1,1,2,4,5,6,7,9,10,12,\\
&& ii)\quad r=-1,1,2,-3,4,7,9,10,12,14,\\
&&iii)\quad r=-1,1,2,-3,-5,9,10,12,14,16\\
&&iv)\quad r=-1,1,-3,-5,-7,10,12,14,16,18.
\end{eqnarray*}
The branch (i) is the principal one and the other three are secondary branches. For the principal branch, the coefficients
of the series (\ref{Laurent}) at non-resonances are
\begin{eqnarray*}
&&v_0=1, \quad v_3=0,\\
&&v_8=\frac{1}{6}v_4^2-\frac{5}{3}v_2^2v_4-\frac{1}{3024}v_2\psi_t-\frac{10}{9}v_2v_6-\frac{1}{36288}\psi_y\\
&&\quad\quad +\frac{5}{18}v_2^4+\frac{1}{9072}v_{1_{t}}, \\
&&v_{11}=\frac{1}{1814400}v_{2_{y}}-\frac{1}{181440}v_{4_{t}}-\frac{16}{45}v_5v_6+\frac{1}{226800}v_2v_{2_{t}}\\
&&\quad\quad-\frac{2}{15}v_2^2v_7-\frac{1}{3}v_4v_7+\frac{1}{75600}v_5\psi_t-\frac{2}{45}v_2^3v_5-\frac{2}{15}v_2v_4v_5
-\frac{4}{9}v_2v_9,
\end{eqnarray*}
where $v_1,v_2,v_4,v_5,v_6,v_7,v_9,v_{10},v_{12}$  are arbitrary functions of the variables $y$ and $t$ and $\phi(x,t,y)=x-\psi(t,y)$ .

\vspace{0.3cm}
\noindent For the second branch, the coefficients of (\ref{Laurent}) at non-resonances are
\begin{eqnarray*}
&& v_0=3, \quad v_3=0, \quad v_5=0, \quad v_6=\frac{1}{10080}\psi_t+3v_4v_2-\frac{1}{3}v_2^3,\\
&&v_8=\frac{1}{133056}\psi_y-5v_2^2v_4-\frac{1}{33264}v_{1_{t}}-\frac{11}{2}v_4^2-\frac{1}{5544}v_2\psi_t+\frac{5}{6}v_2^4,\\
&&v_{11}=\frac{1}{7}v_4v_7-\frac{1}{2540160}v_{2_{y}}-\frac{1}{317520}v_2v_{2_{t}}
+\frac{1}{60480}v_{4_{t}}-\frac{2}{7}v_2^2v_7-\frac{4}{7}v_2v_9,\\
&&v_{13}=-\frac{1}{5364817920}\psi_{tt}-\frac{1}{99792}v_2v_{4_{t}}-\frac{5}{1596672}v_4v_{2_{t}}+\frac{1}{338688}v_2^2v_{2_{t}}\\
&&\quad\quad-\frac{1}{6}v_4v_9-\frac{1}{33264}v_7\psi_t+\frac{25}{126}v_2^3v_7+\frac{5}{42}v_2^2v_9-\frac{15}{14}v_2v_4v_7\\
&&\quad\quad+\frac{1}{6386688}v_{4_{y}}+\frac{1}{7451136}v_2v_{2_{y}},
\end{eqnarray*}
where $v_1,v_2,v_4,v_7,v_9,v_{10},v_{12},v_{14}$ are arbitrary functions of the variables $y$ and $t$ and $\phi(x,t,y)=x-\psi(t,y)$.

\vspace{0.3cm}
\noindent For the third branch, the coefficients of (\ref{Laurent}) at non-resonances are
\begin{eqnarray*}
&& v_0=6, \quad v_3=0, \quad v_4=\frac{1}{5}v_2^2, \quad v_5=0,\\
&&v_6=\frac{2}{15}v_2^3+\frac{1}{110880}\psi_t,\quad v_7=0,\\
&&v_8=\frac{23}{75}v_2^4+\frac{1}{432432}v_{1_{t}}+\frac{1}{72072}v_2\psi_t-\frac{1}{1729728}\psi_y,\\
&&v_{11}=\frac{1}{7257600}v_{2_{y}}-\frac{1}{259200}v_2v_{2_{t}}-v_2v_9,\\
&&v_{13}=\frac{1}{285120}v_2^2v_{2_{t}}+\frac{1}{16094453760}\psi_{tt}+\frac{2}{5}v_2^2v_9-\frac{1}{7983360}v_2v_{2_{y}},\\
&&v_{15}=-\frac{1}{432432}v_9\psi_t-\frac{1}{9}v_2^3v_9-\frac{47}{37065600}v_2^3v_{2_{t}}-\frac{1}{523069747200}v_{1_{tt}}\\
&&\quad\quad-\frac{1}{40236134400}v_2\psi_{tt}+\frac{1}{1046139494400}\psi_{ty}+\frac{47}{1037836800}v_2^2v_{2_{y}},
\end{eqnarray*}
where $v_1,v_2,v_9,v_{10},v_{12},v_{14},v_{16}$ are arbitrary functions of the variables $y$ and $t$ and $\phi(x,t,y)=x-\psi(t,y)$.

\vspace{0.3cm}
\noindent For the fourth branch, the coefficients of (\ref{Laurent}) at non-resonances are
\begin{eqnarray*}
&& v_0=10, \quad v_2=0, \quad v_3=0, \quad v_4=0, \quad v_5=0,\\
&&v_6=\frac{1}{480480}\psi_t,\quad v_7=0,\\
&&v_8=-\frac{1}{25945920}\psi_y+\frac{1}{6486480}v_{1_{t}}, \quad v_9=0,\\
&& v_{11}=0, \quad v_{13}=-\frac{1}{80472268800}\psi_{tt},\\
&&v_{15}=-\frac{1}{3835844812800}\psi_{ty}+\frac{1}{1917922406400}v_{1_{tt}},\\
&&v_{17}=-\frac{1}{1183632113664000}\psi_{yy}+\frac{1}{295908028416000}v_{1_{ty}}-\frac{17}{39916800}v_{10_{t}},
\end{eqnarray*}
where $v_1,v_{10},v_{12},v_{14},v_{16},v_{18}$ are arbitrary functions of the variables $y$ and $t$ and $\phi(x,t,y)=x-\psi(t,y)$.

To sum up, the principal branches of $2+1$ KdV$(2n+4)$ equations for $n=1,2,3$ admit arbitrary functions and the compatibility conditions
at all non-negative integer resonances are satisfied identically. Hence $2+1$ KdV$(2n+4)$ equations for $n=1,2,3$ possess the Painlev\'{e} property.

\section{Conclusion}

 We introduced a new class of nonlinear partial differential
equations, $2+1$ KdV($2n+4$) equations, in $2+1$ dimensions derived
from the KdV equation and its symmetries. We have given 3-soliton
solutions of these equations for $n=1,2,3$. We showed that they
also have the Painlev{\' e} property for $n=1,2,3$. We conjecture that
these equations have 3-soliton solutions and possess the Painlev{\'
e} property for all positive integer $n$.

\vspace{1cm}

\section{Acknowledgment}
 The authors would like to thank  Sergei Sakovich for his critical reading of the manuscript and Willy Hereman
 for providing the Maple package for the Painlev{\'e} analysis. This work is partially supported by the Scientific
and Technological Research Council of Turkey (T\"{U}B\.{I}TAK) and
Turkish Academy of Sciences (T\"{U}BA).


\begin{thebibliography}{99}
\bibitem{kar1} A. Karasu-Kalkanli, A. Karasu, A. Sakovich,
S. Sakovich, R. Turhan, J. Math. Phys. {\bf49}, 073516
(2008),(arXiv: nlin/ 0708.3247.)

\bibitem{kup} B. A. Kupershmidt, Phys. Lett. {\bf A372}, 2634 (2008), (arXiv:
nlin/0709.3848).

\bibitem{kun1} A. Kundu, J. Phys. A: Math. Theor. {\bf 41}, 495201
(2008).

\bibitem{kun2} A. Kundu, J. Math. Phys. {\bf 50} 102702
(2009).

\bibitem{kun3} R. Sahadevan, L. Nalinidevi, A. Kundu, J. Phys. A: Math. Theor. {\bf 42},
115213 (2009).

\bibitem{guha} P. Guha, J. Phys. A: Math. Theor. {\bf 42}, 345201
(2009).

\bibitem{kerst} P. H. M. Kersten, I. S. Krasil'shchik, A.M.
Verbovetsky. and R. Vitolo, Acta Appl. Math.  [arXiv:0812.4902v2].

\bibitem{ram} A. Ramani, B. Grammaticos, R. Willox, Analysis and
Applications, {\bf 6}, 401 (2008).

\bibitem{wang} J. P. Wang, J. Phys. A: Math. Thoer. {\bf 42},
362004 (2009).

\bibitem{zou} R. Zhou, J. Math. Phys. {\bf 50}, 123502 (2009).

\bibitem{Waz} A. M. Wazwaz, Appl. Math. Comput. {\bf 204}, 963
(2008).

\bibitem{her} W. Hereman and A. Nuseir, Mathematics and
Computation in Simulations {\bf 43} (1), 13 (1997).

\bibitem{sak} S. Sakovich,  J. Phys. A: Math. Theor. {\bf 47},
L503, (1994).

\bibitem{hirota} R. Hirota, Phys. Rev. Lett. {\bf 27}, 1192
(1971).

\bibitem{hirota1} R. Hirota, {\bf The Direct Method in Soliton
Theory}, Iewanami Shoten (1992) [Japanese], English Transl.
Cambridge University Press, Cambrdige (2004).

\bibitem{hiet} J. Hietarinta, J. Math. Phys. {\bf 28}, 1732
(1987).

\bibitem{hiet1} J. Hietarinta, Proceedings of the 1991
International Symposium on Symbolic and Algebraic Computation,
ISSAC'91, ed. Stephen Watt (Association for Computing Machinery).

\bibitem{hiet2} B. Grammaticos, A. Ramani, and J. Hietarinta, J.
Math. Phys. 2572 (1990).

\bibitem{Xu-Li} G.Q. Xu, Z.B. Li, Applied Mathematics and Computation {\bf 169}, 1364-1379 (2005).

\bibitem{WTCKruskal}  G.Q. Xu, Z.B. Li, Chin. Phys. Lett. {\bf 20}, 975-978 (2003).

\bibitem{WTC} J. Weiss, M. Tabor, G. Carnevale, J. Math. Phys. {\bf 24}, 522-526 (1983).

\bibitem{Kruskal} M. Jimbo, M.D. Kruskal, T. Miwa, Phys. Lett. A {\bf 92}, 59 (1982).



\end{thebibliography}
\end{document}